\documentclass[10pt,journal]{IEEEtran}
\usepackage{amsmath,amsfonts}
\usepackage{algorithmic}
\usepackage{algorithm}
\usepackage{array}
\usepackage[caption=false,font=normalsize,labelfont=sf,textfont=sf]{subfig}
\usepackage{textcomp}
\usepackage{stfloats}
\usepackage{url}
\usepackage{verbatim}
\usepackage{graphicx}
\usepackage{cite}
\usepackage{multirow}
\begin{document}

\title{Functional Connectivity Methods for EEG-based Biometrics on a Large, Heterogeneous Dataset}

\author{Pradeep Kumar G,
        Utsav Dutta,
      Kanishka Sharma,  ~Ramakrishnan~Angarai Ganesan,% <-this % stops a space
\thanks{Pradeep and Ramakrishnan are with the Department of Electrical Engineering, Indian Institute of Science, Bangalore, India. e-mail: pradeepkg@iisc.ac.in.}
\thanks{Utsav Dutta is with University of Maryland, USA.}
\thanks{Kanishka is with Centre for Neuroscience, Indian Institute of Science.}}

\markboth{Journal of \LaTeX\ Class Files,~Vol.~01, No.~1, June~2022}%
{Shell \MakeLowercase{\textit{et al.}}: A Sample Article Using IEEEtran.cls for IEEE Journals}

% column for its text to clear the IEEEpubid mark. guy 

\maketitle

\begin{abstract}
This study examines the utility of functional connectivity (FC) and graph-based (GB) measures with a support vector machine classifier for use in electroencephalogram (EEG) based biometrics. Although FC-based features have been used in biometric applications, studies assessing the identification algorithms on heterogeneous and large datasets are scarce. This work investigates the performance of FC and GB metrics on a dataset of 184 subjects formed by pooling three datasets recorded under different protocols and acquisition systems. The results demonstrate the higher discriminatory power of FC than GB metrics. The identification accuracy increases with higher frequency EEG bands, indicating the enhanced uniqueness of the neural signatures in beta and gamma bands. Using all the 56 EEG channels common to the three databases, the best identification accuracy of 97.4\% is obtained using phase-locking value (PLV) based measures extracted from the gamma frequency band. Further, we investigate the effect of the length of the analysis epoch to determine the data acquisition time required to obtain satisfactory identification accuracy. When the number of channels is reduced to 21 from 56, there is a marginal reduction of 2.4\% only in the identification accuracy using PLV features in the gamma band. Additional experiments have been conducted to study the effect of the cognitive state of the subject and mismatched train/test conditions on the performance of the system. 
\end{abstract}


\section{Introduction}
In the recent past, functional connectivity (FC) based features derived from electroencephalogram (EEG) have been used for different classification problems. 
%Li et al. \cite{li2019eeg} combined power spectral density, differential entropy and features derived from functional connectivity to classify positive, neutral and negative valence states. Wang \cite{wang2021brain} used different FC measures along with a support vector machine classifier to classify the EEG into one of four emotional states based on valence and arousal. Li et al. \cite{li2020depression} combined functional connectivity matrices with a convolutional neural network to classify EEG from mild depression patients from that of normal subjects. Peng et al. \cite{peng2019depression} used the functional connectivity based phase lag index feature from the resting stage EEG to classify patients with depression from healthy controls. Alba et al. \cite{alba_CNP2016_ADHD} showed that the temporal variability of resting state FC indices are effective in distinguishing patients with attention deficit hyperactivity disorder from control subjects. Lorenzo et al. \cite {lorenzo_FHN2015_schizo} has reported that the resting state EEG source FC characterizes schizophrenic patients from normal subjects and displays differing patterns in the different frequency bands. Wang et al. \cite{wang2021fatigue} used phase lag index features derived from the different EEG frequency bands to detect driving fatigue.

\par Biometric systems have been traditionally built around distinguishing physical characteristics such as fingerprints, iris scans, gait, and speech \cite{jain2007handbook}. In the case of fingerprints and iris scans, it has been shown that synthetic samples can be used to circumvent the identification algorithms used in these systems \cite{adler2015biometric}. Most biometric fingerprint sensors rely on capturing only a fraction of the entire fingerprint for data acquisition and subsequent matching from a database to classify the subject. Researchers at New York University showed that generative adversarial networks could be used to synthetically create fractions of fingerprints, called DeepMasterPrints \cite{bontrager2018deepmasterprints}. These synthetic fingerprints were falsely classified as actual fingerprints 20\% of the time.  The ease with which fingerprints can be collected from surfaces and subsequently replicated on a thin film made of rubber/silicon makes them susceptible to circumvention \cite{yager2004fingerprint,yang2019security}. In biometric systems, iris images are stored in a low-dimensional format as a code in a shared database. Researchers showed that given the code database and corresponding iris scans, it is possible to reverse-engineer the image of the iris and print it onto a contact lens. This method could generate samples which successfully fooled the system 80\% of the time. High-resolution images can be used to recreate a printed eye lens which can be used to bypass security measures \cite{gupta2016survey},\cite{ruiz2008direct}. Although the above biometric modalities are cost-effective and accurate, the possibility of circumvention raises issues in high-security applications.

\par Novel characteristics based on walking gait \cite{goffredo2009self}, speech \cite{fox2007robust}, heart signals \cite{pinto2018evolution}, passphrase keyboard dynamics \cite{bhana2020passphrase}, and ear identification \cite{kumar2012automated} have been proposed as possible biometrics since they are harder to imitate \cite{patel2015cancelable}. EEG systems, which record the electrical activity of different brain areas in the form of a multivariate time series, have been proposed for use as potential biometric systems \cite{chan2018challenges}. With the advent of contactless sensors, it is conceivable to incorporate contactless EEG systems \cite{chi2009non,svard2010design} for subject identification in high-security facilities. EEG data suffers from low signal-to-noise ratio and scale-based shifts in the absolute value of the signals. Hence, this study uses functional connectivity-based metrics which quantify the relation between neural activity observed in different regions of the scalp \cite{bastos2016tutorial}. Recent work on EEG biometrics have primarily been on finding better discriminative features and identification techniques and the use of deep learning frameworks \cite{chan2018challenges}. Commonly used features are Fourier coefficients, autoregressive (AR) model parameters, \cite{liu2013individual,palaniappan2007biometrics} wavelet coefficients, Shannon entropy, and spectral entropy \cite{thomas2016biometric}. Power spectral density (PSD) and AR features have achieved accuracies in the range of 95-99\% \cite{yang2017usability}.

\subsection{Related Works}
Chang et al. \cite{chang2020eeg} combined the directed FC measures with signal complexity measures for person recognition systems with the best accuracy of 90.6\% using delta band and SVM. Rocca et al. \cite{la2014human} implemented spectral coherence measures with Mahalanobis distance-based classifier for subject identification. This study highlighted the increased discriminatory power of modeling the neural activity as interconnected links in the form of connectivity values over the univariate metrics like power spectral density. Further, it showed that the brain's frontal regions reflected unique, subject-specific neural activity due to the influence of genetic factors. Fraschini et al. \cite{fraschini2019robustness} highlighted the robustness of phase-locking value (PLV) and correlation over varying brain states. The study described the potential pitfalls of such metrics, since they are affected by volume conduction and muscle artifacts and proposed using source reconstruction methods to mitigate these effects. The same group \cite{fraschini2018eeg} and reported that the least equal error rate (EER)  for person identification of 5.9\%  is achieved on Dataset-1 considered in the present study using PLV measure rather than the other FC measures in the gamma band. Another study \cite{kong2019eeg} reported increased accuracy with higher frequency bands, namely beta and gamma using PLV measure with a mean accuracy of 99\% evaluated on Dataset-1. Although this study used three datasets, the performance of the identification algorithm was not tested on the combined dataset.

\par Lower-dimensional features can be extracted from the connectivity matrix by modeling the electrodes as nodes in a graph network \cite{farahani2019application}. Node degree, clustering coefficient, and centrality measures are scalar features that represent a graph network by assigning a score to each node. Wang et al. \cite{wang2019convolutional} used convolutional neural networks (CNN) on the connectivity values represented as a graph network. The study achieved 99\% accuracy for two datasets comprising 109 and 59 subjects evaluated separately using PLV combined with the graph CNN approach. However, there was a significant degradation in accuracy when trained and tested across different mental states and the algorithm was not tested on the combined dataset. Fraschini et al. \cite{fraschini2015eeg} used eigenvector centrality and a graph network derived metric by modeling the brain connectivity network using phase lag index (PLI) values as edge weights and achieved an EER of 0.044 (\(\sim\)96\% accuracy) for the identification of 109 subjects. The work highlighted the pitfalls in comparing EEG biometric systems due to the variabilities in the EEG systems, protocols, dataset size, and methodological biases.

\subsection{Contribution of the study}
\par Most EEG biometric studies have used relatively small datasets for validating methods based on FC and graph-based (GB) metrics \cite{yang2017usability}, with all the data used having been collected under identical conditions following the same protocol and with the same EEG system. A recent review on EEG biometric authentication \cite{JalalyBidgoly2020} highlighted that 80\% of studies used datasets of 4 to 50 subjects. To the best of the knowledge of the authors, the study by \cite{chen2019eeg} is the only one with data from 157 subjects combined from heterogeneous datasets. 96\% accuracy was reported in identifying 157 subjects using GLST-CNN method and 93\%, with SVM. Lack of variability in system settings and conditions can lead to optimized results for a particular dataset/data acquisition protocol, which may not be optimal on another dataset, resulting in poor generalization.

\par
\begin{figure*}[ht]
\begin{center}
\includegraphics[width=18cm]{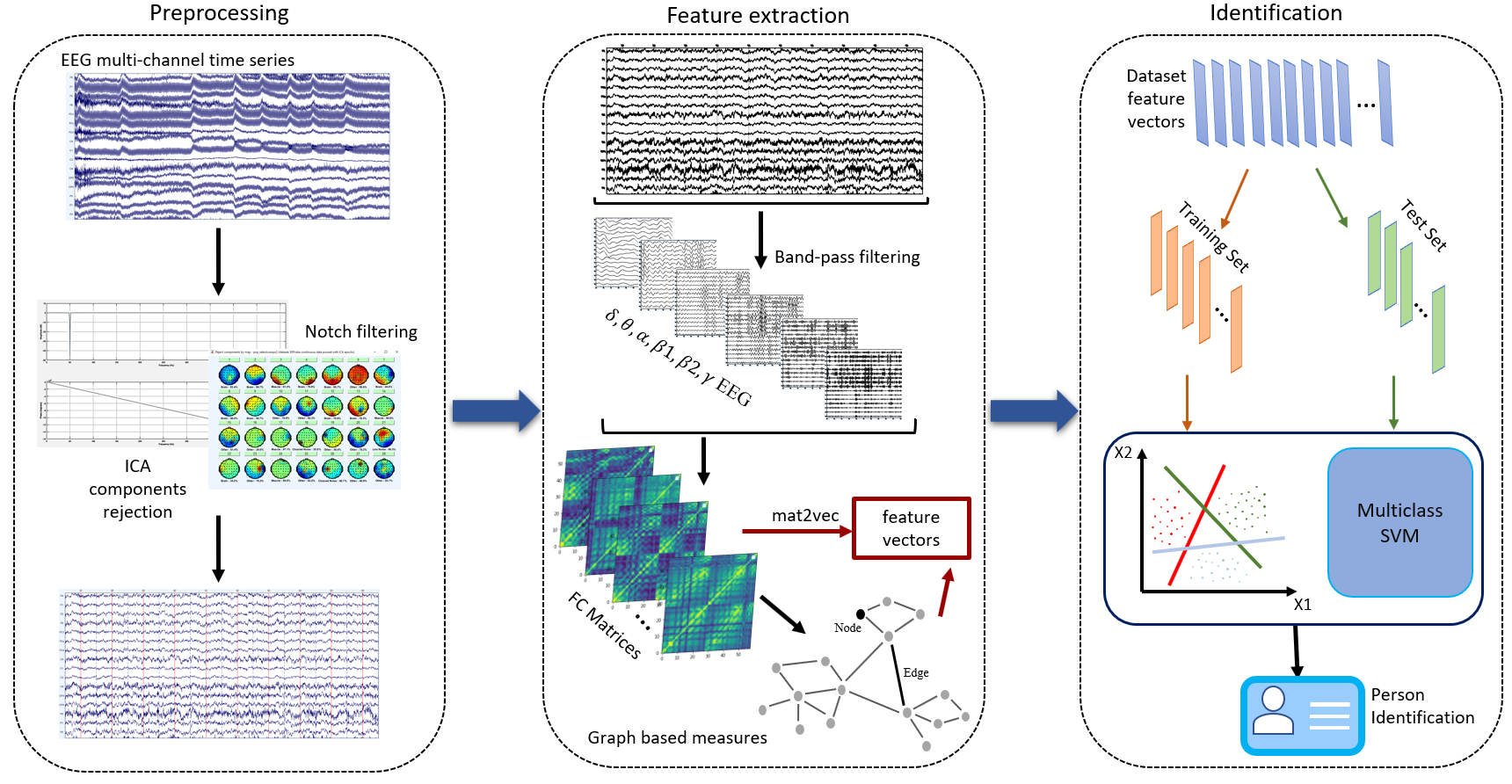}
\end{center}
\caption{Schematic representation of the proposed EEG-based biometric system using functional connectivity or graph-based measures: Raw EEG data is pre-processed and bandpass filtered into different frequency bands. FC and graph network-based measures are extracted from each frequency band, which are then input to the classifier for training, and also testing.}
\label{BlockDiagram1}
\end{figure*}

 Our study presents an EEG-based biometric framework that explores the potential of FC and GB measures on a heterogeneous dataset from 184 subjects. We also examine the potential of the individual frequency bands of EEG for biometrics. Figure ~\ref{BlockDiagram1} presents the proposed methodology. Our work involves a dataset more comprehensive than the ones used in the studies reported in the literature. The data is pooled together from different experimental protocols and recording systems by considering common EEG channels and resampling each dataset to 128 Hz. This increases the generalizability of the technique to data acquired from any EEG system. We use SVM classifier and double k-fold or nested crossvalidation. The study also assesses the performance of two phase-based measures in person identification systems based on EEG. Finally, varying epoch lengths are considered to observe the resulting changes in the identification accuracy.

\renewcommand{\arraystretch}{1.5}
\begin{table*}[ht]
\caption{ Details of the three EEG datasets used for the study. In all the cases, only one minute data from the resting state was used for the experiments. Even though each database contains 64 channels, the combined dataset considers only the 56 channels common to all the databases. All the experiments reported use only these 56 channels. To study the effect of reduced number of channels, experiments have also been conducted using only 21 channels and also using task-based data.}
\label{DatasetDetails}
\small
\begin{center}
\begin{tabular}[c]{l l c c c c}
\hline\hline
\textbf{Dataset} &  \textbf{\shortstack{EEG data \\ collection setup}} & \textbf{\shortstack{\# Channels \\ Recorded\textbackslash Considered }} & \textbf{\shortstack{Original sampling \\ frequency (Hz)}} & \textbf{\# Epochs} & \textbf{\# Subjects} \\
\hline\hline
Dataset-1 & Motor Movement/Imagery & 64\textbackslash 56, 21 & 160 &1635 &109 \\

Dataset-2 & Meditation + Control & 64\textbackslash 56, 21 & 1024 &   300 &  14+6 \\

Dataset-3 & Meditation & 64\textbackslash 56, 21 & 500 &   825 &  55 \\ \hline

Combined & - & -\textbackslash 56, 21 & - & 2760 & 184 \\
\hline\hline
\end{tabular}
\end{center}
\vspace{-0.5cm}
\end{table*}

\section{Materials and Methods}
\subsection{Datasets utilized in the study}
The dataset used for the experiments is pooled from three different recordings to assess the applicability of the derived metrics for data obtained using different systems under distinct conditions and protocols as tabulated in Table ~\ref{DatasetDetails}. Dataset-1 is a publicly available dataset. The authors recorded datasets 2 and 3, and ethical clearance was obtained from \textit{IISc Institute Ethical Committee with IHEC No: 23-24072019}.

\subsubsection{Dataset-1 (Public dataset)}
The PhysioNet Motor Imagery dataset \cite{schalk2004bci2000} \cite{goldberger2000physiobank} (Dataset-1) consists of EEG recorded from 109 subjects using the 64-channel BCI2000 system at a sampling rate of 160 Hz. It has 14 blocks related to motor imagery visualization tasks.

\subsubsection{Dataset-2 (Recorded by authors)}
This has EEG recorded from 20 subjects (age 30-52 years, mean 43.9, SD=4.0) with a 64-electrode Waveguard cap and ANT Neuro mylab system at a sampling rate of 1024 Hz. 14 of the 20 subjects are long-term meditators who practiced two types of meditation after separate baseline segments \cite{ganesan2020characterization}. The remaining subjects are controls from whom only baseline data of duration 3 minutes has been collected.

\subsubsection{Dataset-3 (Recorded by authors)}
Dataset-3 contains EEG data recorded from 55 long-term meditators (age: 42.0 ± 10.1 years, range: 25-59; meditation experience: 17.5 ± 10.8 years, range: 4 -43; 14 females) with a 64-electrode Waveguard cap and ANT Neuro mylab system at a sampling rate of 500 Hz. The protocol involves a meditation segment with eyes-open and eyes-closed baseline segments.

For this study, only 1-minute eyes-open baseline data is considered for each of the above three datasets. It is split into 4-second long, non-overlapping epochs  \cite{sharma2020brain}, resulting in 15 epochs per subject.
 
\subsubsection{The Combined Dataset}
This is formed by pooling the 1-minute, resting state, eyes-open segments of Datasets-1, 2, and 3. Since the sampling rates are different, each dataset is first independently resampled to 128 Hz, both for independent and combined dataset analysis. 56 channels common across the datasets are selected for feature extraction. The same 56 channels have been used for the person identification experiments involving the individual and the combined datasets.

\subsection{Pre-processing and bandpass filtering}
The datasets are preprocessed using a standard pipeline, which includes notch filtering to eliminate the line noise at 50 Hz, band-pass filtering from 0.5 to 45 Hz, decomposition by independent component analysis, and manual artifact rejection using EEGLAB \cite{delorme2004eeglab}. To study the effectiveness of different frequency components for the task in hand, the data is divided into six frequency bands using Butterworth band-pass filters \cite{widmann2015digital}. The features defined in Secs. 2.3 and 2.4 are extracted for each band of frequencies listed in Table ~\ref{FrequencyBands}.

\subsection{Functional Connectivity (FC) Features}
FC metrics assess the relationship between the EEG data from different regions of the scalp \cite{bastos2016tutorial}. Traditional FC measures are bivariate, although extensions to multivariate cases have been proposed \cite{peng2019depression} recently. Scalar FC metrics are computed between every pair of electrodes over a pre-defined epoch. Since the connectivity matrix is symmetric, only its upper triangular elements are concatenated into a row vector of dimension \(N_{ch} (N_{ch} -1)/2\), where \(N_{ch}\) is the number of channels used for the experiments. The FC measures used in this study are explained in the following subsections.

\renewcommand{\arraystretch}{1.4}
\begin{table}[ht]
\captionsetup{justification=centering}
\centering
\caption{EEG bands and their frequency ranges used in the study.}
\label{FrequencyBands}
\scriptsize
\begin{center}
\begin{tabular}{|c||c||c|c|c|c|c|}
\hline
\textbf{EEG band} & Delta & Theta & Alpha & Beta1 & Beta2 & Gamma  \\
\hline
\textbf{Freq range (Hz)} & 0.5 - 4  & 4 - 8  & 8 - 12   & 12 - 20 & 20 - 30  & 30 - 45  \\
\hline
\end{tabular}
\end{center}
%\vspace{-0.5cm}
\end{table}

\subsubsection{Pearson's Correlation (COR)}
This metric quantifies the linear relationship between two signals, \(x_m\) and \(x_n\). A value of zero indicates no linear association, and a value of one shows a perfect relationship \cite{cohen2014analyzing}.
\begin{equation}
\rho_{m,n}=\frac{\sum_{k=0}^{M-1} (x_{m,k}-\bar{x}_m)(x_{n,k}-\bar{x}_n)}{\sigma(x_m)\sigma(x_n)}  
\end{equation}
\(x_n\), \(\bar{x}_n\), and \(\sigma(x_n)\) are the EEG data from the \(n^{th}\) channel, its mean, and standard deviation, respectively; \(M\) is the number of samples in each analysis interval or epoch. With the use of all the 56 common channels, we get a total of $\binom{56}{2} = 1540$ COR values, which form the feature vector.

\subsubsection{Phase Locking Value (PLV)}
The phase locking value represents the average of the phase differences between a pair of time series. Higher values indicate robust clustering of the phase differences around a particular value on the polar plot \cite{cohen2014analyzing}. In our study, the instantaneous phase \(\phi_m(k)\) of the time series $x_m$ is obtained using Hilbert transform for each electrode \(m\). PLV between the respective epochs of length \(M\) samples of two time series $x_m$ and $x_n$ is defined as, \begin{equation}
    PLV_{m,n}=\left\lvert \frac{1}{M}\sum_{k=0}^{M-1}e^{j(\phi_{m}(k)-\phi_{n}(k))}\right\rvert
\end{equation}

\subsubsection{Phase Lag Index (PLI)}
PLI is the measure of the average ‘leading’ or ‘lagging’ nature of one time series with respect to another throughout a particular time segment \cite{cohen2014analyzing}. The instantaneous phase is obtained as mentioned in section 2.3.2 and PLI between the corresponding epochs of two time series $x_m$ and $x_n$ is defined as,
\begin{equation}
    PLI_{m,n}=\left\lvert\frac{1}{M}\sum_{k=0}^{M-1}sgn(\phi_{m}(k)-\phi_{n}(k))\right\rvert
\end{equation}

The \(sgn\) function thresholds the differences in phase values to \(-1\), \(1\) or \(0\) depending on the sign of the difference.

\[sgn(x)=\begin{cases}
-1 & \text{if $x<0$} \\
 0 & \text{if $x=0$} \\
 1 & \text{if $x>0$} \\
\end{cases} \] 

\subsection{Features based on Graph Network Analysis}
A graph $G$ is defined as an ordered pair of disjoint sets of nodes and edges: \(G=(V,E)\), where $V$ is the set of nodes and \(E \subseteq \{\{x,y\}\mid x,y \in V\) and \(x \neq y\}\) is the set of edges. In graph network analysis, the connectivity matrix is expressed as an undirected graph \cite{bullmore2011brain}, consisting of a non-zero valued edge connection between every pair of nodes. Each electrode is treated as a node, and the value of a chosen connectivity metric between the signals of a pair of electrodes is assigned as the weight of the edge connecting the two nodes. Four graph network metrics are derived from the values of each of the three connectivity metrics assigned as the edge weights. Each of the graph-based metric results in a $N_{ch}$-dimensional vector whose elements are the ‘scores’ assigned to the different channels included. Thus, the graph network analysis approach results in a low-dimensional representation of the EEG data.

\subsubsection{Node degree (ND)}
This captures information on the network's local structure by measuring the depth of a node's ties to all its neighbors. Node degree is defined as,
\begin{equation}
    d_m=\sum_{\substack{n=1; n \neq m}}^{N_{el}} w_{m,n}
\end{equation}
where \(d_m\) is the degree of the m-th node, n refers to all the nodes of the graph other than m and \(w_{m,n}\) is the edge weight connecting nodes \(m\) and \(n\). Suppose ND is computed using the PLV values as the edge weights. With $56$ channels, we get a feature vector of dimension 56 comprising the ND values of all the channels common to the three databases used.

\subsubsection{Eigenvector Centrality (EC)}
The eigen-decomposition of the \(N_{ch}\times N_{ch}\) connectivity matrix is evaluated as,
\begin{equation}
    C_{N_{ch}\times N_{ch}}=PDP^{-1}
\end{equation}
where, \(D\) is the diagonal matrix of the eigen values, and columns of \(P\) contain the eigenvectors of the matrix \(C\). The first column of $P$ is the dominant eigenvector corresponding to the maximum variance in the data and is used as a feature. It is denoted by $[e_1 e_2 ... e_{N_{ch}}]^T$. The \(i^{th}\) element $e_i$ of this vector contains the contribution of the \(i^{th}\) channel to the eigenvector \cite{fraschini2015eeg}. Thus, EC results in a feature vector of dimension \(N_{ch}\).

\subsubsection{Betweenness Centrality (BC)}
BC of a particular node $u$ quantifies the degree to which it is part of the interactions between other nodes in the network \cite{makarov2018betweenness}. It is measured as,
\begin{equation}
    c_b(u)=\sum_{\substack{m,n \in V \\m \neq n \neq u}} \frac{\sigma(m,n\mid u)}{\sigma(m,n)}
\end{equation}
where, \(\sigma(m,n)\) is the number of shortest weighted paths between the pair of nodes \(m\) and \(n\); \(\sigma(m,n\mid u)\) is the subset of them passing through node \(u\). A higher value of \(c_b\) indicates a higher fraction of paths passing through the given node. We compute the betweenness of every node using the algorithm proposed by Bentert et al. \cite{bentert2018adaptive}. It associates the higher information transfer for long paths with higher weights (or connectivity). The weight of each path is equal to the sum of the weights encountered on the path.

\begin{figure*}[ht]
\begin{center}
\includegraphics[width=14cm]{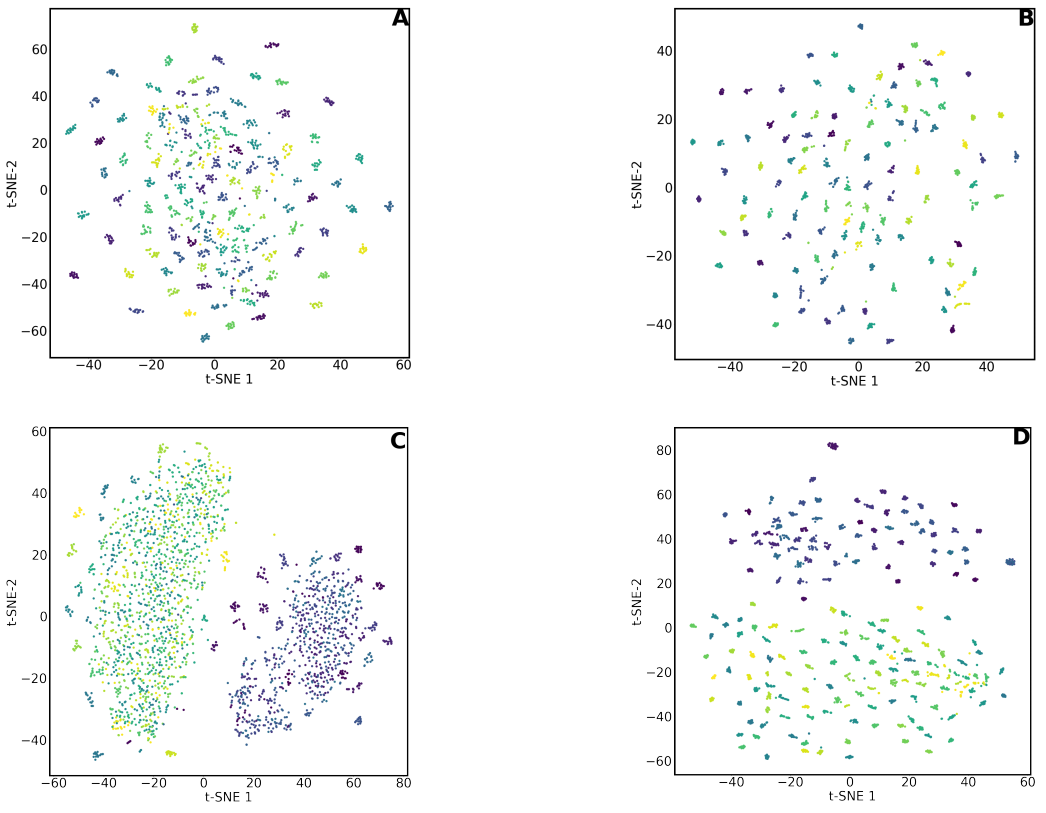}
\end{center}
\caption{Lower dimensional visualization using t-SNE of PLV adjacency matrix extracted from \textbf{(A)} alpha band, \textbf{(B)} gamma band of Dataset-1 (109 subjects) and \textbf{(C)} alpha band, \textbf{(D)} gamma band of combined dataset (184 subjects). Each subject’s samples are identified by a unique shade of colour, chosen as distinct hues on a colour palette. The gamma band signal leads to good separation of the subjects in the feature space.}
\label{t-SNE plots}
\end{figure*}

\subsubsection{Clustering Coefficient (CC)}
This captures the degree to which nodes in the graph cluster locally \cite{demuru2020comparison}. A higher value of CC indicates increased information transfer in local networks around the electrode. The CC of a node $u$ is defined as:
\begin{equation}
    c_c(u)=\frac{1}{d_{u}(d_{u}-1)} \sum_{\substack{m,n\in V\\ m \neq n \neq u}} (\hat{w}_{u,m}\hat{w}_{u,n}\hat{w}_{m,n})^\frac{1}{3}
\end{equation}
where \(d_{u}\) is the node degree, and \(\hat{w}_{u,m}={w_{u,m}}/{w_{max}}\) , where \(w_{u,m}\) is the connectivity value between nodes \(u\) and \(m\) using the chosen FC metric, and $w_{max}$ is the highest valued element in the functional connectivity matrix.

\subsection{Discriminability of the different features}
In order to visualize the ability of the different features defined in Secs. II.C and II.D to discriminate between the subjects, a dimensionality reduction technique is applied to them. t-distributed stochastic neighbor embedding (t-SNE) \cite{van2008visualizing} is a stochastic, non-deterministic method used for reducing the dimensionality of any data while retaining the relative ‘distance’ between points across dimensions. t-SNE minimizes the KL-divergence between the probability distributions of the data in the high and low dimensional spaces. We use t-SNE to reduce the feature representation to two dimensions for visualization purposes. The t-SNE plots in Fig. ~\ref{t-SNE plots} show that the clustering of data from different subjects is more robust in higher frequencies (gamma band) across the FC and GB metrics. This further motivated the study’s objective of building a model for subject identification. Since the t-SNE algorithm uses Gaussian kernels for building probability density estimates, algorithms using similar kernel expressions, such as radial basis function (RBF) kernels in support vector machines, may perform well for such types of datasets. Hence we use the SVM classifier with the RBF kernel.

\subsection{Subject Identification Experiments}
Each of the features defined in Sec.s 2.3 and 2.4 in its vectorized form of connectivity matrix is assessed individually for each of the four datasets. This study uses multi-class SVMs \cite{meyer2003support} which are capable of dealing with high dimensional data with minimal tuning. The SVM used in this case employs a 'one-vs-rest' approach where a different binary classification model is trained for each target class. RBF kernel is used, assigning a weight to each point based on its relative distance to points belonging to the same class.

\par A double k-fold or nested cross-validation approach \cite{stone1974cross} is used to assess the algorithm's performance. The dataset is split into k1 = 10 folds, of which (k1-1) folds form the training set, and one fold is held out as the unseen test set. The training set is then divided into k2 = 3 folds, of which one fold is held out as a validation set while the remaining two folds are combined to train the model. For each of the k1 folds, the best performing hyperparameter set is used to estimate the performance on the unseen test set. For each set of features, the SVM model is optimized using grid search of the hyperparameters C (0.1, 1, 10, 100) and gamma (1, 0.1, 0.01, 0.001).

\subsection{Experiments using lower number of EEG channels}
To study the effect of the number of channels employed on the biometric identification accuracy, additional experiments have been conducted using only the subset of 21 EEG channels corresponding to the 10-20 electrode system. If very good performance can be obtained with a few channels, it will facilitate real-life deployment of EEG-based biometrics.

\subsection{Experiments to study the effect of cognitive state}
EEG signal is sensitive to the cognitive state of the person. Different mental activities involve distinct cortical networks, which result in distinct functional connectivity configurations. Thus the biometric recognition performance of a EEG-based system may depend on the cognitive state of the subject. To test this, we conducted experiments by training and testing our system using data corresponding to the task, rather than the resting baseline. In our case, the tasks vary across the datasets. Dataset 1 corresponds to motor imagery; datasets 2 and 3 involve EEG from distinct meditative states. We conducted these experiments for both 56 and 21-channel data and compared their performance. Only the PLV feature is used since it is the best performing feature in the initial experiments involving the resting state data from all the 56 channels.

\subsection{Experiments with mismatched train/test conditions}
The performance of any biometric system is sensitive to the conditions of the enrolment and test data being comparable, and accuracy generally degrades when the test data corresponds to conditions different from the enrolment time. In our case, all the original experiments were conducted under matching conditions of resting state. However, to study the robustness of the system under mismatched train/test conditions, we also performed experiments where the enrolment uses the resting state EEG, whereas the testing uses task-based EEG and vice versa.

\begin{table*}
    \caption{System performance on resting state EEG data from all the 56 channels common to the three databases, for each frequency band. Identification accuracies (in \%) on all the datasets using different FC metrics. The best accuracy obtained for each dataset is shown in bold. PLV feature extracted from the gamma band performs the best on all the datasets.}
\label{FCMetricAccuracy}
    %\tiny
    \centering
    \begin{tabular}{|c|c|c|c|c|c|c|c|}
        \hline
        \textbf{Data} & \textbf{FC metric} & \textbf{Delta} & \textbf{Theta} & \textbf{Alpha} & \textbf{Beta1} & \textbf{Beta2} & \textbf{Gamma} \\ \hline\hline
        \multirow{3}{*} {\shortstack{Dataset-1 \\ (109 subjects)}} & PLI & 12 $\pm$ 1.4 & 21 $\pm$ 1 & 47 $\pm$ 5.1 & 79 $\pm$ 1.3 & 95.4 $\pm$ 1.1 & 99.1 $\pm$ 0.8 \\ \cline{2-8}
         & PLV & 76 $\pm$ 3 & 76 $\pm$ 2.3 & 76 $\pm$ 3.8 & 93$\pm$1.0 & 96 $\pm$ 0.9 & \textbf{99.4$\pm$ 0.7} \\ \cline{2-8}
         & COR & 70$\pm$2.1 & 69$\pm$3.1 & 69$\pm$ 4.6 & 91.4$\pm$1.6 & 97$\pm$0.6 & 97.8$\pm$0.7 \\  \hline\hline
        \multirow{3}{*}{\shortstack{Dataset-2 \\ (20 subjects)}} & PLI & 9$\pm$3.5 & 18$\pm$7.6 & 28$\pm$5.6 & 19$\pm$3.7 & 32$\pm$4.7 & 46$\pm$8.4 \\ \cline{2-8}
         & PLV & 89$\pm$2 & 91$\pm$3 & 88$\pm$2.6 & \textbf{93$\pm$1} & 92$\pm$2 & \textbf{93$\pm$0.1} \\ \cline{2-8}
         & COR & 88$\pm$4.9 & 88$\pm$4.7 & 87$\pm$2.9 & 91$\pm$1.6 & 92$\pm$0.02 & \textbf{93$\pm$0.2} \\ \hline\hline
        \multirow{3}{*}{\shortstack{Dataset-3 \\ (55 subjects)}} & PLI & 5$\pm$2.9 & 10$\pm$1.8 & 5$\pm$2.9 & 10$\pm$3.3 & 8$\pm$2.6 & 18$\pm$2.7 \\ \cline{2-8}
         & PLV & 73$\pm$3.3 & 68$\pm$4.4 & 70$\pm$4.8 & 89$\pm$2.2 & 96$\pm$1.2 & \textbf{97$\pm$0.4} \\ \cline{2-8}
         & COR & 73$\pm$3.2 & 66$\pm$5.2 & 68$\pm$4.9 & 87$\pm$2.4 & 95$\pm$1.1 & 96$\pm$1.0 \\  \hline\hline
        \multirow{3}{*}{\shortstack{Combined \\ (184 subjects)}} & PLI & 8$\pm$1.8 & 15$\pm$3.4 & 27$\pm$2.4 & 53$\pm$1.3 & 61$\pm$1.6 & 68$\pm$1.5 \\ \cline{2-8}
         & PLV & 73$\pm$3.3 & 71$\pm$3.6 & 68$\pm$2.9 & 91$\pm$1.2 & 96$\pm$1.0 & \textbf{97.4$\pm$0.8} \\ \cline{2-8}
         & COR & 67$\pm$2.8 & 60$\pm$3.1 & 60$\pm$3.2 & 86$\pm$1.9 & 94$\pm$1.1 & 95.7$\pm$0.6 \\  \hline
    \end{tabular}
\end{table*}

\renewcommand{\arraystretch}{1.2}
\begin{table*}[ht]
% \makegapedcells
\small
    \centering
    \caption{System performance on resting state EEG using graph-based features. Identification accuracies on all the datasets for the 3 highest performing GB features derived from 3 different FC metrics across the frequency bands using the common 56 EEG channels. For all the datasets, the node degree feature extracted from the PLV of gamma band performs the best.}
    \label{GBmetricsAccuracy}
    \begin{tabular}{|c|c|c|c|c|}
    \hline
        \textbf{Dataset} & \textbf{FC metric} & \textbf{GB metric} & \textbf{Band} & \textbf{Accuracy} \\ \hline\hline
         \multirow{3}{*}{\shortstack{Dataset-1 \\ (109 subjects)}} & PLV & Node degree & Gamma & \textbf{97\%} \\ \cline{2-5}
         & COR & Node degree & Gamma & 96\% \\ \cline{2-5}
         & PLI & Node degree & Gamma & 93\% \\ \hline\hline
         \multirow{3}{*}{\shortstack{Dataset-2 \\ (20 subjects)}} & PLV & Node degree & Gamma & \textbf{95\%} \\ \cline{2-5}
         & COR & Node degree & Gamma & 95\% \\ \cline{2-5}
         & PLV & Clustering coefficient & Gamma & 94\% \\ \hline\hline
         \multirow{3}{*}{\shortstack{Dataset-3 \\ (55 subjects)}} & PLV & Node degree & Gamma & \textbf{92\%} \\ \cline{2-5}
         & PLV & Node degree & Beta2 & 83\% \\ \cline{2-5}
         & PLV & Clustering coefficient & Gamma & 79\% \\ \hline\hline
         \multirow{3}{*}{\shortstack{Combined \\ (184 subjects)}} & PLV & Node degree & Gamma & \textbf{91\%} \\ \cline{2-5}
         & COR & Node degree & Gamma & 89\% \\ \cline{2-5}
         & PLV & Node degree & Beta2 & 83\% \\ \hline
    \end{tabular}
\end{table*}

\section{Results}
\subsection{Performance using different features from different bands}
Performance on each of the datasets using different FC and GB metrics are given in Tables ~\ref{FCMetricAccuracy} and ~\ref{GBmetricsAccuracy}, respectively, while using all the 56 channels. The reported values are average double k-fold cross-validated identification accuracies with the standard error. The best accuracies obtained are 99.4\%, 95\%, 97\% and 97.4\% for the datasets 1, 2, 3 and the combined one, respectively. For all the datasets except dataset-2, the optimum feature for which the highest accuracy is obtained is the bivariate FC metric PLV derived from the gamma band. Reducing the connectivity matrices to graph representations (Table ~\ref{GBmetricsAccuracy}) significantly reduces the discriminatory power of the algorithm, except for Dataset-2, for which the node degree feature gives the best performance of 95\% for both PLV and COR. However, the maximum accuracy returned by any FC metric for Dataset-2 is only 93\%. The obtained identification accuracy values are low, when the FC or GB measures are extracted from the complete EEG signal (full bandwidth). Hence, these results are not reported.

\par Further, it is observed that the optimal FC features are from higher frequency bands beta1, beta2, and gamma, indicating higher discriminatory neural activity at higher frequencies. Across all the datasets, FC measure PLV-gamma uniformly displays strong predictive capability across different sampling frequencies, subjects, data acquisition protocols, and EEG systems, thus highlighting its potential general-purpose use with different EEG datasets. In the literature, there is only one other study by \cite{chen2019eeg}, which deals with a large number of subjects. The results of the above work are compared with those of the present study in Table ~\ref{studyComparison}. Even with a deep neural network, they could achieve a performance of only 96.3\%, whereas our method has achieved a recognition rate 1.1\% above this value on a database with 27 more subjects.

\subsection{Effect of epoch length}
We analyze the effect of epoch length on the identification accuracy using the best performing feature of FC metric PLV-gamma on the combined dataset. As the analysis segment considered is 60 seconds long, the number of epochs and the length per epoch are related as follows:
\begin{equation}
   \# of epochs = {60}/{epoch length} 
\end{equation}
\(\# of epochs\) directly determines the number of feature samples available to the classifier. Higher the number of samples, higher the identification performance due to better generalization. In contrast, increased epoch length is also expected to increase the identification accuracy by providing more robust estimates for the connectivity metrics as measured over a more extended time. However, as the epoch length increases, the number of feature samples reduces. Since the length of the available data is fixed, the number of epochs and the epoch length cannot be increased simultaneously. Thus we aim to find the optimal epoch length which would maximize the algorithm’s performance. We evaluate the classifier’s performance for epoch lengths of 2 to 6 sec as given in Table ~\ref{EffectofEpochlength}. Epoch length of 4 seconds results in the highest accuracy, and hence the analysis has been carried out with 4-sec epochs.

\renewcommand{\arraystretch}{1}
\begin{table}
\small
    \centering
        \caption{Effect of the epoch length on person identification performance on the combined dataset (184 subjects) using the best performing feature, PLV extracted from the gamma band.}

\label{EffectofEpochlength}
\begin{tabular} {|c|c| }
    \hline
        \textbf{Epoch length} & \textbf{Accuracy} \\ \hline\hline
        2s & 96.64\% \\ \hline
        3s & 97.21\% \\ \hline
        \textbf{4s} & \textbf{97.35\%} \\ \hline
        5s & 96.82\% \\ \hline
        6s & 96.42\% \\ \hline
\end{tabular}
\vspace{-0.5cm}
\end{table}

\begin{table*}
    \caption{Performance using reduced number of channels. Person identification accuracies (in \%) on all the 4 datasets using different FC metrics derived from 21 electrodes (10-20 system) and each frequency band. The best accuracy obtained for each dataset is shown in bold. PLV feature extracted from the gamma band performs the best on all the datasets.}
\label{FCMetric-21channels}
    %\tiny
    \small
    \centering
    \begin{tabular}{|c|c|c|c|c|c|c|c|}
        \hline
        \textbf{Data} & \textbf{FC metric} & \textbf{Delta} & \textbf{Theta} & \textbf{Alpha} & \textbf{Beta1} & \textbf{Beta2} & \textbf{Gamma} \\ \hline\hline
        \multirow{3}{*} {\shortstack{Dataset-1 \\ (109 subjects)}} & PLI & 9 $\pm$ 2 & 7 $\pm$ 2 & 14 $\pm$ 3 & 38 $\pm$ 3 & 65 $\pm$ 4 & 89 $\pm$ 2 \\ \cline{2-8}
         & PLV & 47 $\pm$ 3 & 46 $\pm$ 4 & 46 $\pm$ 4 & 77$\pm$5 & 87 $\pm$ 4 & \textbf{96$\pm$ 2} \\ \cline{2-8}
         & COR & 45$\pm$3 & 43$\pm$4 & 44$\pm$ 3 & 75$\pm$4 & 86$\pm$4 & 94$\pm$3 \\  \hline\hline
        \multirow{3}{*}{\shortstack{Dataset-2 \\ (20 subjects)}} & PLI & 10$\pm$4 & 16$\pm$8 & 24$\pm$7 & 15$\pm$6 & 19$\pm$5 & 30$\pm$8 \\ \cline{2-8}
         & PLV & 83$\pm$6 & 89$\pm$3 & 86$\pm$4 & {92$\pm$2} & \textbf{93$\pm$1} & \textbf{93$\pm$0} \\ \cline{2-8}
         & COR & 85$\pm$5 & 87$\pm$5 & 83$\pm$5 & 88$\pm$3 & \textbf{93$\pm$1} & \textbf{93$\pm$1} \\ \hline\hline
        \multirow{3}{*}{\shortstack{Dataset-3 \\ (55 subjects)}} & PLI & 4$\pm$2 & 6$\pm$2 & 1$\pm$3 & 8$\pm$2 & 6$\pm$3 & 12$\pm$4 \\ \cline{2-8}
         & PLV & 55$\pm$8 & 58$\pm$4 & 59$\pm$4 & 87$\pm$4 & 95$\pm$2 & \textbf{96$\pm$1} \\ \cline{2-8}
         & COR & 57$\pm$6 & 61$\pm$4 & 62$\pm$4 & 83$\pm$3 & 94$\pm$2 & \textbf{96$\pm$1} \\  \hline\hline
        \multirow{3}{*}{\shortstack{Combined \\ (184 subjects)}} & PLI & 5$\pm$1 & 6$\pm$1 & 13$\pm$2 & 26$\pm$2 & 43$\pm$2 & 59$\pm$1 \\ \cline{2-8}
         & PLV & 51$\pm$3 & 53$\pm$2 & 54$\pm$3 & 81$\pm$2 & 89$\pm$2 & \textbf{95$\pm$1} \\ \cline{2-8}
         & COR & 50$\pm$3 & 49$\pm$2 & 49$\pm$2 & 76$\pm$2 & 88$\pm$1 & 93$\pm$2 \\  \hline
    \end{tabular}
\end{table*}

\subsection{Results of additional experiments}
Table ~\ref{FCMetric-21channels} lists the performance obtained using the FC metrics PLI, PLV and COR derived considering only the classical 21-channel subset of the data. Comparing the results in Table ~\ref{FCMetricAccuracy}, we see that the performance degrades heavily for all the features derived from the lower frequency bands. Gamma band gives the best accuracy of 95\% for the PLV feature, which is only 2.4\% less than the corresponding figure for the complete 56-channel data. Thus there is promise for EEG biometrics even with lower number of channels.

\begin{figure}[ht]
\begin{center}
\includegraphics[width=9cm]{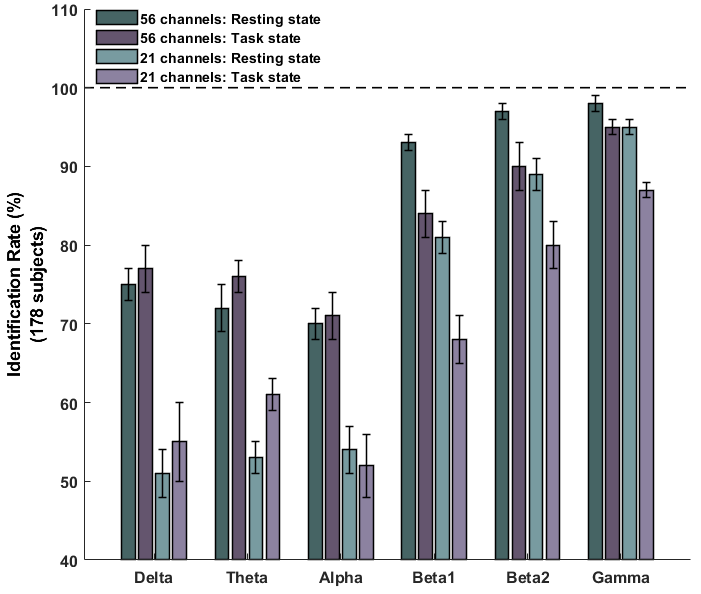}
\end{center}
\caption{System performance using task-based EEG data. Crossvalidation accuracies for each frequency band from 56 or 21 channels in resting and task state conditions. Since 6 of the 20 subjects in dataset2 are controls, for whom there is no meditation data, for fair comparison, the results are shown only for the remaining 178 subjects. The results shown are for the best feature of PLV. In the gamma band, which entails the best accuracy, performance degrades only by about 2\% with the use of task-based data.}
\label{BarChartMatched}
\vspace{-0.5cm}
\end{figure}

In the second set of additional experiments, the classifier is both trained and tested on different non-overlapping segments of task-based (meditation or motor imagery) data. Fig. ~\ref{BarChartMatched} gives the bar chart comparing the identification accuracies obtained using the task-based data with the performance using resting data - both for 56 and 21-channel data. Since 6 of the 20 subjects in dataset2 are control subjects, for whom there is no meditation data, for fair comparison, the results of all the experiments are shown only for the remaining 178 subjects. For delta, theta, and alpha bands, the performance with task-based data is superior by a few percentage. With beta1, beta2 and gamma bands, the resting state data entails better accuracies. Using the gamma band, identification accuracies of 97 and 94\% are obtained with resting and task-based data, respectively. Using the classical subset of 21 channels, the performance is dismally low for low frequency bands, but gradually improves with increasing frequencies. Using 21-channel, resting state, gamma-band data, the performance is lower than that of 56-channel by  only 2.5\%. However, with the task-based data, the performance comes down by nearly 9\% with the use of only the 10-20 electrodes.

\begin{figure*}[ht]
\begin{center}
\includegraphics[width=14cm]{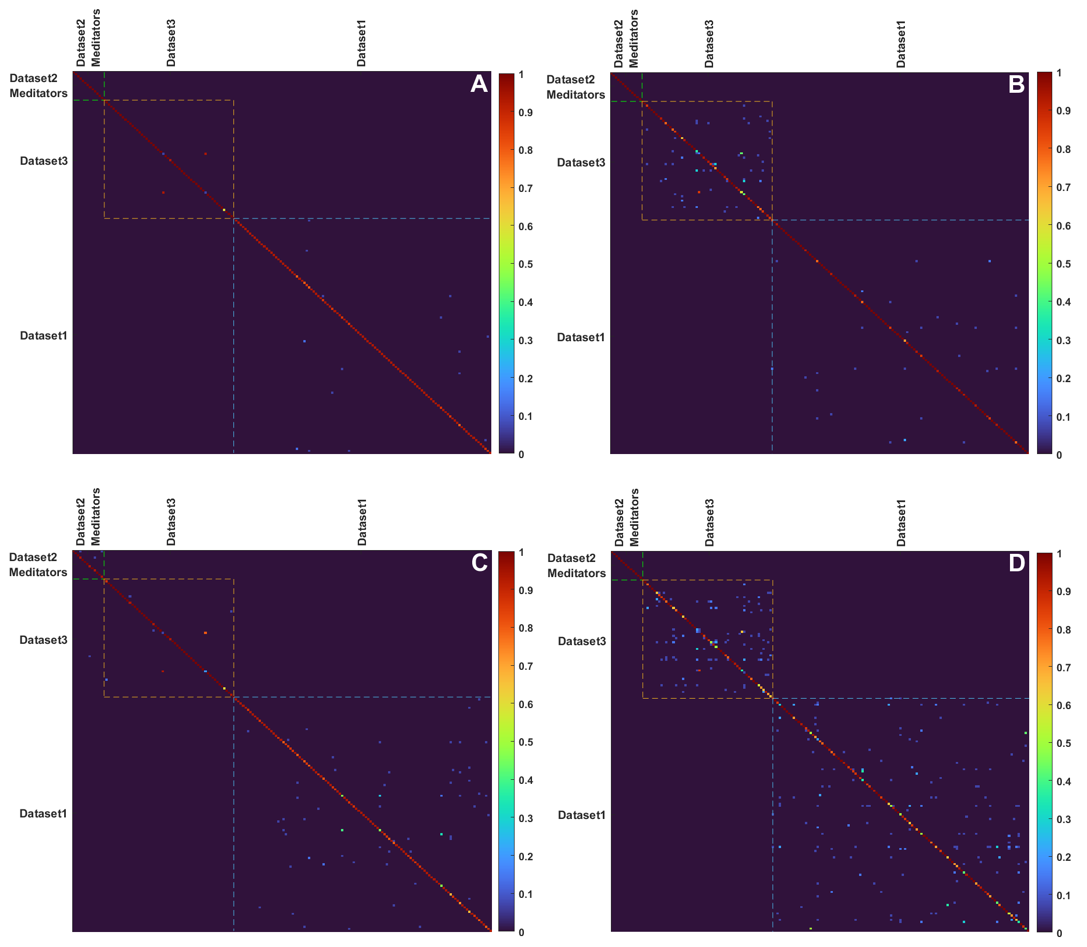}% This is a *.eps file
\end{center}
\caption{Confusion matrices (shown as images) using PLV feature from the gamma band. For \textbf{(A)} resting-state, and \textbf{(B)} task-based data with 56 channels (178 subjects). For \textbf{(C)} resting, and \textbf{(D)} task data with 21 channels. The dotted lines with different colors represent subjects from different datasets. The confusions increase with the use of lower number of channels and/or the task-based data. Thus, maximum confusions arise with the use of 21-channel task-based data.}
\label{ConfusionMatrices}
\end{figure*}

\begin{figure}[ht]
\begin{center}
\includegraphics[width=09cm]{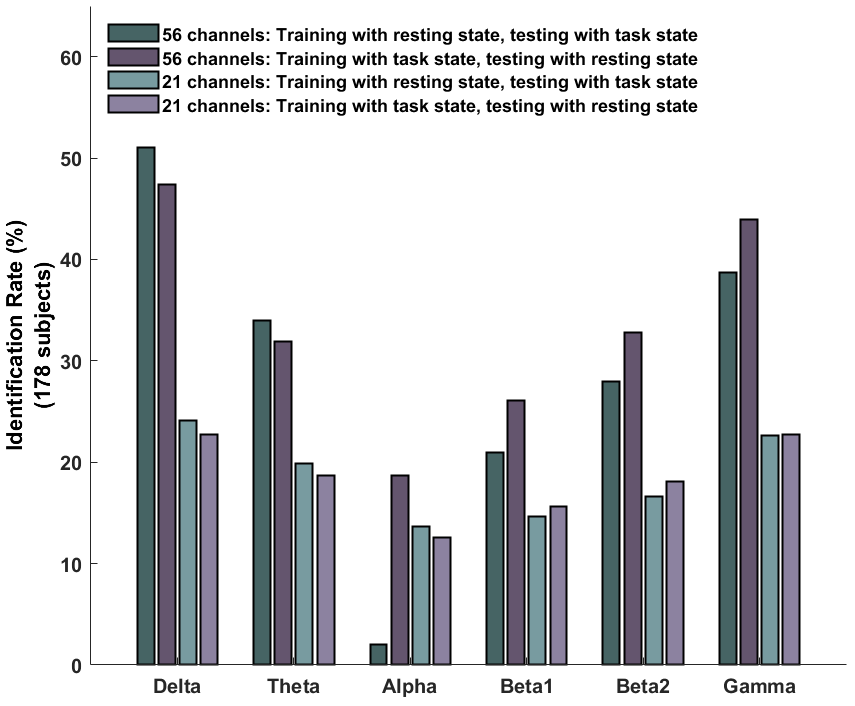}
\end{center}
\caption{Performance with mismatched train/test conditions employing high and low number of channels. Identification rates obtained with 56 and 21 EEG channels for different frequency bands.The training and test data have been recorded during different states (resting versus task or vice versa). Clearly, there is significant reduction in the biometric performance when the testing conditions differ from the training conditions.}
\label{Mismatch56-21}
\vspace{-0.5cm}
\end{figure}

Fig. ~\ref{ConfusionMatrices} shows the confusion matrices for four distinct cases. With 178 subjects, it is difficult to show a complete confusion table using numbers. Hence, we have shown the confusion tables as color-coded images, where different hues represent different levels of confusion among the subjects. Fig. ~\ref{ConfusionMatrices}C shows that confusion increases when less number of EEG channels are used for person identification. Figs. ~\ref{ConfusionMatrices}B and D show that the confusion significantly increases with task-based data irrespective of whether 56 or 21 channels are used. %The maximum number of confusions that any subject’s data has had is ? (this indicates the assignment of a subject’s data to other subjects) and the minimum number of confusions that any subject’s data has had is zero.

The results of experiments with mismatched train/test data are illustrated by bar charts in Fig. ~\ref{Mismatch56-21}. Compared to the matched conditions, there is a significant reduction in the identification accuracy when the enrolment data is from the resting state and the test data is from a task or vice versa. For the high frequency bands, there is marginally less reduction in the accuracy when the enrolment uses task-based EEG data and the test data is from the resting state. This trend reverses for low frequency bands. Thus, interestingly, the resting data - delta band results in the maximum accuracy of about 51\% in the mismatched conditions, whereas it is only 39\% for the resting state - gamma band. The accuracy values are lower than 25\% for all the bands with 21-channel data under the mismatched conditions. Since the performance in mismatched conditions are very low, we have not shown the confusion matrices in this case.

\section{Discussion}
The present study's primary objective is to assess the utility of EEG for person identification systems on heterogenous datasets using functional connectivity. The FC methods used are Pearson's correlation, phase locking value, and phase lag index. It is observed that PLV is a robust discriminatory feature for the combined dataset comprising three different datasets, highlighting its robustness. Accuracy is higher using higher frequency bands due to better clustering of subjects as shown by Fig. ~\ref{t-SNE plots}, with competitive accuracies obtained from beta and gamma bands. Performance of the PLV measure over other phase-based measures in higher frequency bands is consistent with previous studies \cite{fraschini2019robustness,kong2019eeg,fraschini2018eeg}. 
Following this, we assess the trade-off between the number of training samples and the epoch length. Since these two are mutually conflicting, it is expected that for a fixed length of data, an intermediate value of epoch length would provide the optimal number of training samples to result in the best performance and a sufficient epoch length to obtain strong estimates of the connectivity features. This analysis is conducted on the combined dataset using the PLV-gamma feature, and the algorithm performs optimally at an epoch length of four seconds. Higher accuracy is observed with FC measures than features such as autoregressive and power spectral density features (see Table ~\ref{studyComparison}).

Traditionally, the alpha band was believed to be the main rhythm characterizing the resting state. However, recently, the resting state gamma band has gained attention in different EEG applications. Kavitha et. al \cite{thomas2018eeg} reported higher accuracy using a vector of gamma power values of all the channels at the resting state as the feature. Crobe et. al \cite{crobe2016minimum}  obtained the least equal error rate (EER) using the gamma band while analyzing subject-specific EEG traits. Fraschini et. al \cite{fraschini2015eeg} reported lowest EER with resting state, gamma band, EEG data for biometric application using the eigenvector centrality measure. Gamma PSD was significantly high in a default mode network versus dorsal attention network comparison. Ventral attention and language networks are majorly modulated by gamma during resting state \cite{mantini2007electrophysiological}. With the small network synchronization in high frequency band and its role in self-referential processing, a direction is emerging to use features derived from gamma band as distinctive markers for individual identification based on EEG.

\section{Conclusion}
The methodology of using functional connectivity-based metrics with an SVM classifier is proposed for subject identification. The  results achieved on heterogenous datasets totalling 184 subjects are competitive to studies reported on EEG biometrics in the recent past, although the accuracy reported by such studies are on a single dataset recorded under identical conditions. The dataset1 that we have used is from Physionet, which was recorded employing an EEG system and for tasks distinct from the ones we used to record the other datasets. The results are only marginally lower when the number of channels are reduced from 56 to 21. Thus, our results show that PLV feature derived from the gamma band is robust for biometric identification and works effectively independent of the EEG system or the recording conditions, as long as the type of data used for enrolment and testing are matched. It is evident that univariate measures such as AR coefficients and power spectral densities have been outperformed by models using connectivity-based measures that capture additional information on the interaction between the different brain regions. The results of the current study are consistent with those of other studies comparing FC measures, which also have obtained optimal results in high-frequency bands with phase-based measures quantified by PLV.

In continuation with this study, one can explore different feature sets, such as multivariate functional connectivity measures. Exploring the use of CNNs or spatio-temporal CNN-LSTM networks with raw EEG data for automated feature extraction may result in better discriminatory power. However, \cite{chen2019eeg} has achieved a performance of only 96.3\% using their GSLT-CNN framework on the raw EEG signal. Thus, our features, together with the SVM classifier, have performed well, comparable to the best results in the literature on similar large datasets.

Large scale studies with datasets collected over more extended periods under altered mental states are necessary to validate the use of EEG as a general-purpose biometric. The advent of contactless sensors would increase the viability of EEG biometric systems for high-security purposes. Being able to obtain reliable recording accounting for movement, varying head sizes, etc. poses its own challenges for EEG data collection for biometrics. Mathematical models to account for such measurement-related variability are essential for obtaining a repeatable reading from these devices.

\par Since EEG provides better protection against replication or mimicking than the traditional biometric systems, it is conceivable that it will attract greater attention in high-security applications. Building systems that extract adequate discriminatory information using significantly fewer electrodes combined with low-cost EEG acquisition devices may enable such technologies to be deployed for more mundane applications.

\renewcommand{\arraystretch}{1.7}
\begin{table*}[ht]
\centering
\caption{Performance comparison with a similar study for individual identification on a heterogeneous EEG dataset. Our results on a dataset of higher number of subjects are comparable or better than those of Chen et al. for any feature and/or classifier.}
\label{studyComparison}
%\tiny
% \begin{center}
\begin{tabular}{|l| c| c| c |c |l| c|}
\hline
\textbf{Study} &  \textbf{\# Subjects} &\textbf{\# Channels} & \textbf{Frequency band} & \textbf{Classifier} & \textbf{         Feature set} & \textbf{Accuracy}\\
\hline\hline
Chen et al., 2019 & 157 & 28 & 0.1-32 Hz & SVM & {\shortstack{Autoregressive coefficients}}  & 86.3\% \\
\hline
Chen et al., 2019 & 157 & 28 & 0.1-32 Hz & SVM & {\shortstack{Power spectral density}} & 92.9\% \\
\hline
Chen et al., 2019 & 157 & 28 & 0.1-32 Hz & GSLT-CNN & {\shortstack{Raw EEG data}} & 96.3\% \\
\hline
Proposed study & 184 & 21 & {\shortstack{Gamma (30-45 Hz)}} & SVM & {\shortstack{Functional connectivity measures}} & \textbf{95\%} \\ \hline 
Proposed study & 184 & 56 & {\shortstack{Gamma (30-45 Hz)}} & SVM & {\shortstack{Functional connectivity measures}} & \textbf{97.4\%} \\ \hline 
\end{tabular}
% \end{center}
\end{table*}

\bibliographystyle{IEEEtran}
\bibliography{biometricpaper.bib}

\begin{comment}

\end{comment}

%\newpage

%vspace{11pt}

\begin{IEEEbiography}[{\includegraphics[width=1in,height=1.25in,clip,keepaspectratio]{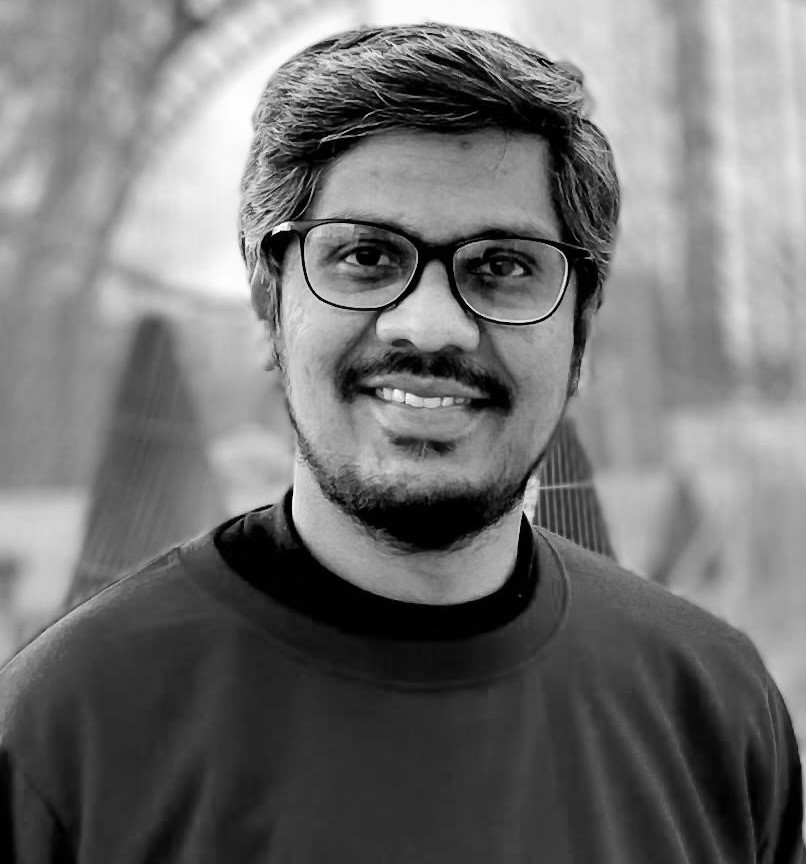}}]{Pradeep Kumar G} received his BE in electronics and communication engineering and M.Tech in digital electronics and communication engineering from Visveswaraya Technological University, India. He is pursuing the Ph.D degree in system science and signal processing at Department of Electrical Engineering, Indian Institute of Science (IISc). He studies the neural mechanisms and correlates during various Non-Ordinary states of consciousness (Meditation, Hypnosis, Trance) using EEG and the effects of these states on cognitive functions and classification of meditative states from resting states. His research interests include cognitive neuroscience and neural signal processing.
\end{IEEEbiography}
\vspace{-1.0cm}
\begin{IEEEbiography}[{\includegraphics[width=1in,height=1.25in,clip,keepaspectratio]{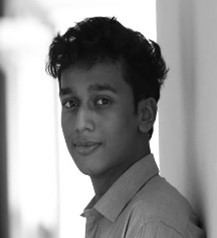}}]{Utsav Dutta} is a research assistant at the Medical Intelligence and Language Engineering laboratory at the Indian Institute of Science, Bangalore. He received his bachelor’s degree in Mechanical Engineering from the Indian Institute of Technology, Madras. His research interests include machine learning, statistical modelling, and biomedical data science. 
\end{IEEEbiography}
\vspace{-1.0cm}
\begin{IEEEbiography}[{\includegraphics[width=1in,height=1.25in,clip,keepaspectratio]{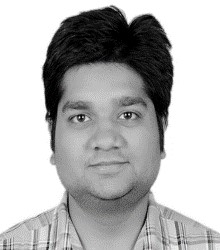}}]{Kanishka Sharma} received the B.Sc. degree in life-sciences from Dr. B.R. Ambedkar University, Agra in 2008 and the M.Sc. degree in Neuroscience from Jiwaji University, Gwalior in 2010 and Ph.D. in Cognitive Neuroscience from University of Delhi, New Delhi in 2019.
\par He is a Post-doctoral fellow funded by the Cognitive Science Research Initiative, Dept. of S \& T, Govt. of India, in the Department of Electrical Engineering, IISc. From 2012 to 2017, he was a Research Scholar with the Biomedical engineering lab at Institute of Nuclear Medicine and Allied Science, Delhi. His research interests include psychophysiological experiments to explore human cognitive indices, meditation, and contemplative practices for cognitive enhancement. Dr. Kanishka’s awards and honors include the Junior and Senior research Fellowship (Defence R\&D Organization, Ministry of Defence, Govt. of India), the ICMR travel grant, travel award by International Brain Research Organization, the DST International travel award by Department of S \& T, Govt. of India. His paper won the best paper award in IEEE Bangalore Humanitarian Technology Conference held at Bengaluru and was highlighted in national print and TV media.
\end{IEEEbiography}
\vspace{-1.0cm}
\begin{IEEEbiography}[{\includegraphics[width=1in,height=1.25in,clip,keepaspectratio]{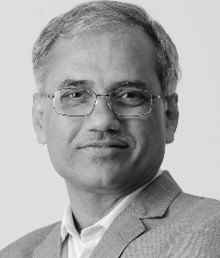}}]{Ramakrishnan A G}is a Professor of Electrical Engineering at the IISc. He obtained his Ph. D. and M Tech from the Indian Institute of Technology Madras and completed his B.E. (Hons) at PSG College of Technology. He heads the Medical Intelligence and Language Engineering Lab at the IISc, Bangalore. He is a Fellow of Indian National Academy of Engineering and an Associate Editor of Frontiers in Neuroscience.
\par He was a Senior Research Scientist at HP Labs, Bangalore. His research interests include Neural Signal Processing, Machine Learning, and Speech Synthesis. He received the Manthan Awards 2014 and 2015, for his projects 'Gift of New Abilities' and 'Madhura - The Gift of Voice'. He is also one of the Founder Directors of RaGaVeRa Indic Technologies Pvt Ltd, a deep tech startup incubated by IISc.
\end{IEEEbiography}
\vfill

\end{document}